\documentclass[aps, pra, twocolumn, notitlepage, 10pt, superscriptaddress
]{revtex4-2}

\usepackage{amssymb,latexsym,amsmath}
\usepackage{booktabs}
\usepackage{bm}
\usepackage{amssymb,amsmath}
\usepackage{graphicx,xcolor}
\usepackage[titletoc, title]{appendix}
\usepackage[caption=false,lofdepth,lotdepth]{subfig}
\usepackage{hyperref} 
\usepackage{soul}
\hypersetup{%
	pdfpagemode=FullScreen,
	pdfstartpage=1,
	pdfmenubar=true,
	pdftoolbar=true,
	colorlinks = true,
	linkcolor=blue,
	citecolor=blue,
	urlcolor=blue,
	bookmarksopen=false
}

\begin{document}

\title{Controlled acoustic-driven vortex transport in coupled superfluid rings}

\author{A.~Chaika}
\affiliation{Department of Physics, Taras Shevchenko National University of Kyiv, 64/13, Volodymyrska Street, Kyiv 01601, Ukraine}
\author{A.~O.~Oliinyk}
\affiliation{Department of Physics, Taras Shevchenko National University of Kyiv, 64/13, Volodymyrska Street, Kyiv 01601, Ukraine}

\author{I.~V.~Yatsuta}
\affiliation{Department of Condensed Matter Physics, Weizmann Institute of Science, Rehovot 7610001, Israel}
\author{M. Edwards}
\affiliation{Department of Physics, Georgia Southern University, Statesboro, Georgia 30460-8031, USA}

\author{N.~P.~Proukakis}
\affiliation{Joint Quantum Centre Durham-Newcastle, School of Mathematics, Statistics and Physics, Newcastle University, Newcastle upon Tyne, NE1 7RU, United Kingdom}

\author{T.~Bland}
\affiliation{Mathematical Physics, LTH, Lund University, Post Office Box 118, S-22100 Lund, Sweden}

\author{A.~I.~Yakimenko}
\affiliation{Department of Physics, Taras Shevchenko National University of Kyiv, 64/13, Volodymyrska Street, Kyiv 01601, Ukraine}
\affiliation{Dipartimento di Fisica e Astronomia Galileo Galilei,
Universit\'a di Padova,
Via Marzolo 8, 35131 Padova, Italy}

	\date{\today}

\date{\today}

\begin{abstract}

Atomtronic quantum sensors based on trapped superfluids offer a promising platform for high-precision inertial measurements where the dynamics of quantized vortices can serve as sensitive probes of external forces. We analytically investigate persistent current oscillations between two density-coupled Bose–Einstein condensate rings of equal radius and show that the vortex dynamics is governed by low-energy acoustic excitations circulating through the condensate bulk. The oscillation frequency and damping rate are quantitatively predicted by a simplified hydrodynamic model, in agreement with Bogoliubov–de Gennes analysis and Gross–Pitaevskii simulations. We identify the critical dissipation separating persistent oscillations from overdamped vortex localization. Furthermore, we demonstrate that periodic modulation of the inter-ring barrier at resonant frequencies enables controlled vortex transfer even when the condensates are well separated in density. These results clarify the role of collective hydrodynamic modes in circulation transfer and establish a framework for employing vortex dynamics in atomtronic quantum technologies.
\end{abstract}

\maketitle

\section{Introduction}
Persistent currents in ring-shaped Bose–Einstein condensates (BECs) provide a rich platform for exploring quantum hydrodynamics, phase coherence, and topological excitations in superfluid systems. Such flows, quantized due to the single-valuedness of the macroscopic wavefunction, are central to the field of atomtronics, a growing area that seeks to engineer coherent matter-wave circuits analogous to electronic devices \cite{Amico2021,polo2024persistent}. Systems supporting controllable persistent currents are particularly promising for quantum technologies such as interferometry \cite{Ryu2013,Eckel2014} and precision sensing \cite{woffinden2023viability,pelegri2018quantum}.

A growing body of work has explored various mechanisms for generating, stabilizing, and manipulating persistent currents in atomtronic circuits. Quantized circulation in strongly interacting fermionic superfluids across the BEC-BardeenCooper-Schrieffer (BCS) crossover has been directly detected via interferometric readout in a ring trap \cite{cai2022persistent}, and their metastability in defect-free geometries has been characterized over long timescales \cite{xhani2025stability}. In contrast, decay triggered by obstacles has been linked to vortex-induced phase slips \cite{xhani2023decay}. Theoretical proposals have outlined atomtronic platforms based on multi-component quantum fluids and engineered topologies to support coherent transport \cite{polo2024perspective,polo2024persistent}. Rydberg-ring arrays have demonstrated chiral excitation currents imprinted via Raman dressing, robust to dephasing and disorder \cite{perciavalle2023controlled}. Persistent flows can also be stabilized by distributing the phase winding across multiple junctions in a Josephson necklace, increasing the critical circulation without reducing coherence \cite{pezze2024stabilizing}. 

Recent advances have highlighted the role of compound geometries, such as coupled or stacked condensate rings \cite{Oliinyk2019,oliinyk2019tunneling,escriva2019tunneling,nicolau2020orbital,su2013kibble,schubert2025josephson,bazhan2022josephson,perez2022coherent,PhysRevB.100.205109,PhysRevA.107.023305,adeniji2024double}, in enabling coherent circulation transfer between topologically distinct superfluid components. { Using a different set-up of density-connected co-planar double-ring Bose–Einstein condensates~\cite{bland2020persistent}, we previously demonstrated that vortex-induced current oscillations arise naturally in such systems when linked by a tunable barrier~\cite{Bland_2022}: in such cases, angular momentum is exchanged via hydrodynamic coupling through the inter-ring low-density region, rather than quantum tunneling~\cite{perez2022coherent}.} 
This framework was later extended to include acceleration-driven dynamics, revealing additional control mechanisms and the influence of dissipation on circulation evolution \cite{chaika2024acceleration}. Related studies in vertically stacked toroidal condensates showed that asymmetric preparation can generate rotational Josephson vortices that propagate across the junction \cite{bazhan2022josephson}. Furthermore, circulation transfer driven by sound-like excitations was observed in coplanar ring configurations subjected to asymmetric acceleration \cite{borysenko2025acceleration}, reinforcing the significance of collective hydrodynamic modes in coupled superfluid systems.

In this work, we show that the persistent current oscillations previously reported can be understood as the manifestation of low-energy acoustic normal modes circulating in the double-ring system. Our new approach, which is based on bulk condensate physics -- as opposed to the earlier ghost-vortex picture \cite{Bland_2022} -- not only gives new physical insight into the nature of persistent current oscillation in coupled quantum circuits, but also exhibits quantitative agreement with numerical simulations without the need for parameter tuning. Using linearized hydrodynamic theory and Bogoliubov-de Gennes analysis, we derive accurate predictions for the mode structure, frequency spectrum, and damping behavior. We note that our study not only demonstrates the collective excitation within the system, as it was widely studied in the ring~\cite{Eller2020, kumar2016minimally, cozzini2006vortex, xhani2023decay}, junction~\cite{momme2019collective, Ryu2013}, or dumbbell geometries \cite{Eckel2016,Gauthier2019,lee2013analogs}, but explicitly identifies acoustic excitations with the persistent current transport of distinct vortices, which was not demonstrated previously. The acoustic picture not only explains the emergence of beating and decay in the vortex dynamics but also enables quantitative understanding of circulation transfer in a non-inertial, accelerating frame. Furthermore, we demonstrate that periodic modulation of the inter-ring barrier at resonant frequencies induces controlled phase slips and circulation exchange, even for well-separated rings, via selective excitation of normal modes.

The paper is organized as follows. Section II introduces the model and identifies the regime of persistent current oscillations. Section III analyzes the excitation spectrum under dissipation and acceleration. In Sec. IV, we examine resonant vortex transfer induced by periodic coupling modulation and discuss the experimental relevance of our findings. Section V concludes with a summary and outlook.

\section{Persistent current oscillations}
\subsection{Model and trap geometry}
To describe the behavior of weakly interacting degenerate atoms near equilibrium,  with weak dissipation, we employ the quasi-two-dimensional Gross-Pitaevskii equation, with phenomenological dissipation, widely used in the literature \cite{ choi1998phenomenological,tsubota2002vortex,chaika2024acceleration}
\begin{equation} \label{eq:dGPE2D}
		(i-\gamma)\hbar \frac{\partial \psi }{\partial t} = \bigg(-\frac{\hbar ^{2}}{2 M}
		\nabla ^{2}+V_{\text{ext}}+M \boldsymbol{a} \cdot \boldsymbol{r}+g|\psi |^{2} -\mu_{\rm{2D}}\bigg) \psi\,. 
\end{equation} 
Here, $g$ is the effective two-dimensional two-body local interaction coupling strength, $V_{\text{ext}}$ is an external confining potential, $0 \leq \gamma \ll 1$ is a dissipation factor, and $\mu_{\rm{2D}}$ is the $\rm{2D}$ chemical potential. This equation is formulated in a non-inertial (accelerating) frame, where the term $M \boldsymbol{a} \cdot \boldsymbol{r}$ represents the inertial potential arising from the acceleration.
Its presence takes into account cases where the system is moving under constant acceleration $\boldsymbol{a}$, which we will cover in this work. Adopting the accelerating frame perspective, as opposed to modeling acceleration via a time-dependent potential in the laboratory frame, is essential for a consistent description of relaxation, as it implicitly assumes equilibration with a co-moving thermal cloud. This avoids the unphysical scenario in which damping arises from friction between the {\em accelerating}
condensate and a {\em stationary} thermal background, which would otherwise result from treating acceleration as an external force acting on a static environment. By contrast, starting from a laboratory-frame with a time-dependent potential and transforming the Gross-Pitaevskii equation (GPE) to the accelerated frame generates additional $\gamma$-dependent terms beyond the inertial force, corresponding to an effective relative motion between condensate and reservoir. Therefore, the two approaches are not equivalent, and without an explicit dynamical model for the thermal cloud (beyond the phenomenological dissipative GPE) these extra terms cannot be treated in a controlled manner (see Ref.~\cite{chaika2024acceleration} for a detailed discussion).

We note that the use of the phenomenological dissipative Gross–Pitaevskii equation (GPE) is well justified in this context. In our previous work \cite{Bland_2022}, we benchmarked this approach against more advanced finite-temperature descriptions~\cite{ZNG6,nick_book} -- more specifically against the stochastic projected GPE~\cite{Blakie:2008vka} and the kinetic Zaremba-Nikuni-Griffin~\cite{griffin_nikuni_book_09} model, and found that all methods produced qualitatively consistent results for the oscillation dynamics and damping trends. In the case of a weakly-accelerated system, we assume that the acceleration is constant during the whole protocol so the reservoir and condensate have relaxed to near-equilibrium and the dissipative GPE is applicable. The damped GPE thus provides a reliable and efficient effective model for capturing the dominant features of vortex-mediated transport considered here, even though it does not reproduce mode-dependent Landau damping in full detail.

We proceed using the equal-size double-ring geometry previously used in works~\cite {bland2020persistent,Bland_2022,chaika2024acceleration}. 
The condensate is confined in a double-ring potential of two equal-sized parts  forming the ‘8’-shaped geometry shown in Fig.~\ref{fig: scheme}(a). The trapping potential 
\begin{align}
    V_\text{ext}(\boldsymbol{r},\,t) = V_\text{d}(\boldsymbol{r}) + V_\text{b}(\boldsymbol{r},\,t)
\end{align}
is composed of the static double-ring potential, with rings of equal radius $R$,
\begin{align} \label{eq: trap static}
   V_\text{d}(\boldsymbol{r})=\frac{M \omega_r^2}{2} \min \left((|\boldsymbol{r}+R\boldsymbol{n}|-R)^2,(|\boldsymbol{r}-R \boldsymbol{n}|-R)^2\right) 
\end{align}
and a time-dependent repulsive barrier controlling the connectivity between the two tori via
\begin{align} \label{eq: trap barrier}
     V_\text{b}(\boldsymbol{r},\,t)=V_0(t)\Theta(R-|\boldsymbol{r}\cdot\boldsymbol{n}|)e^{-[\boldsymbol{r}\times\boldsymbol{n}]^{2}/2l_b^{2}}\,.
\end{align}
This barrier is elongated along the axis connecting the centers of the rings and can be implemented via a blue-detuned laser beam. Here, $\boldsymbol{r}=(x,y)$, $\boldsymbol{n}=\left(\cos\theta,\sin\theta\right)$, and $\theta$ is the angle between the direction of acceleration and the line which connects the centers of the rings. At time $t=\Delta t$, the barrier amplitude $V_0(t)$ is linearly increased from zero. When the barrier amplitude exceeds the local chemical potential, it fully depletes the density along its length, merging the two central holes into a single void, as illustrated in Fig.~\ref{fig: scheme}(b). This configuration, referred to as the open-gate regime, effectively transforms the system from two density-linked toroidal condensates into a single, topologically connected toroid, facilitating the transfer of the quantum vortex as discussed in Ref.~\cite{Bland_2022}.

We consider experimentally realistic parameters of the trap, inspired by the experiment of Ref.~\cite{PhysRevLett.110.025302}, but with atom number $N=10^6$. This corresponds to $^{23}$Na atoms with $a_{s}=2.75$~nm, and we also choose $\omega_r = 2\pi\times134$ Hz, $R=22.6~\mu$m, and a potential width of $l_b=3.45~\mu$m. The effective two-dimensional interaction coupling is $g= \sqrt{8 \pi} \hbar^2 a_s/(Ml_z)$, where $l_z=\sqrt{\hbar/M \omega_z}$, and  $\omega_z=2 \pi \times 550$ Hz.  The density is normalized to the total atom number $\iint \,|\psi(x,y,t)|^2\text{d}x\text{d}y =N$, while the chemical potential $\mu_\text{2D}$, in our simulation, is adjusted at each time step, to provide total atom number conservation, similar to that found in Ref. \cite{tsubota2002vortex}.

\subsection{Regimes of persistent current transport}
\begin{figure*}[!ht] 
   \centering   
   \includegraphics[width=1.8\columnwidth]{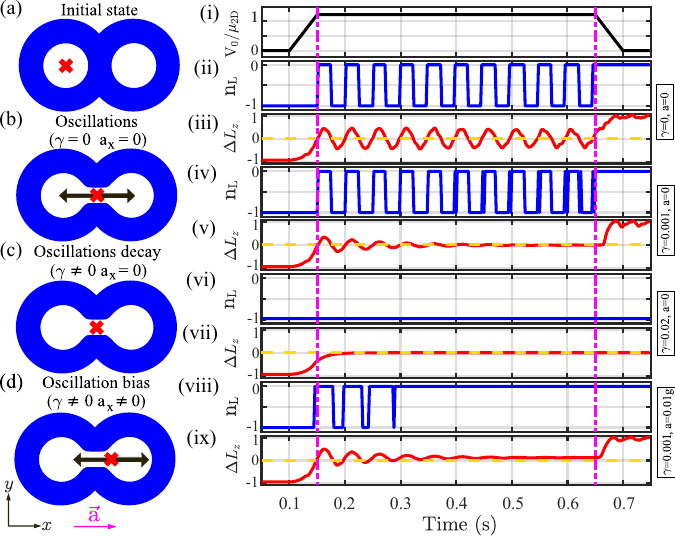}%
\caption{Schematic representation of possible persistent current oscillation regimes. Left column: (a) The prepared initial state with a closed gate and an anti-vortex in the left ring. (b) The state with an open gate, demonstrating the anti-vortex's free oscillation between rings. (c) The open-gate state, which exhibits decaying oscillations due to dissipation. This results in vortex pinning at the center of the system. (d) Biased oscillation due to the presence of acceleration, which shifts the anti-vortex equilibrium position to the right ring. The right column (i) shows the dynamics of the barrier amplitude, where the vertical dot-dashed magenta lines represent the open-gate part of the protocol. The other parts on the right are examples of dynamics for the winding number of the left ring ($\rm{n_L}$) and the angular momentum per particle difference between the rings ($\Delta L_z$) of the corresponding regimes: (ii) and (iii) correspond to the conservative regime (b); (iv) and (v) correspond to the weakly dissipative regime (b-c); (vi) and (vii) correspond to the highly dissipative, overdamped regime (c); and (viii) and (ix) correspond to the biased regime with weak dissipation (d). Yellow dashed lines indicate zero angular momentum difference.}
\label{fig: scheme}
\end{figure*}

The overall setup follows our previous works \cite{Bland_2022, chaika2024acceleration}; their main results are briefly summarized in this section and schematically illustrated in Fig.~\ref{fig: scheme}. 

The initial state [Fig.~\ref{fig: scheme}(a)] is obtained via the imaginary time evolution of an imprinted vortex in a given ring, within a static double-ring potential [Eq.~\eqref{eq: trap static}] at fixed acceleration. During the first $0.05\, \rm{s}$ of real-time evolution, the system evolves under strong dissipation $\gamma=0.02$, to suppress residual excitations and obtain a pure stationary state. 
Thus, after this relaxation, the subsequent dynamics is determined by the metastable persistent-current state (its winding number and stationary density profile), while the specific vortex-preparation method affects only short-lived transients as has been previously explicitly studied in the double-ring geometry by numerical quenched growth dynamics in the context of the stochastic projected GPE \cite{bland2020persistent}.

The subsequent evolution proceeds under the dissipation rate specified in the figure. After a brief closed-system evolution, at time $\Delta t=0.1, \rm{s}$, the barrier amplitude \eqref{eq: trap barrier} [Fig.~\ref{fig: scheme}(i)] begins ramping up to $V_0=1.2 \mu_{\rm{2D}}$, 
transforming the system topology into a torus: henceforth, we will refer to this as the ``open-gate'' state.
(Note that, unless otherwise specified, this work will use the particular value $V_0=1.2 \mu_{\rm{2D}}$ throughout -- with the 
effects of other values discussed in \cite{Bland_2022}.)

For the conservative case, this ``open gate'' allows the vortex to transfer between rings, which leads to indefinite oscillations of the persistent current, as shown in Fig.~\ref{fig: scheme}(b). To detect the vortex position at the open gate stage, we dynamically tracked the following quantities: the winding phase around the left ring, denoted by $\rm{n_L}$, thus $\rm{n_L}=+1(-1)$ means that a vortex(anti-vortex) currently is in the left ring, while $\rm{n_L}=0$ means absence of vorticity there; and the angular momentum per particle difference between the rings. The latter quantity is defined as $\Delta L_z=\langle{L_{z,L}}\rangle-\langle{L_{z,R}}\rangle$, where
\begin{equation*}
    \langle{L_{z,\{L,R\}}}\rangle=\frac{i \hbar}{N_{\{L,R\}}} \iint_{\mathcal{R}}  \, \psi^* \left(y \frac{\partial}{\partial x}-(x \pm R) \frac{\partial}{\partial y}\right) \psi\, \text{d}x \text{d}y.
\end{equation*}
Here, $N_{L/R}$ is the particle number in the left/right ring, respectively, and $\mathcal{R}$ denotes the respective integration region of the left/right part of the dimer. However, as our numerous time-dependent simulations show, both quantities are in agreement, despite producing different values: discrete for the winding number, and continuous for angular momentum, which can be seen in the dynamics of $\rm{n_L}$ and $\Delta L_z$ in Fig.~\ref{fig: scheme}. Therefore, we can establish that phase accumulation in the inter-ring region is rather irrelevant, as it does not significantly contribute to angular momentum, due to negligible density there. Thus, the phase essentially accumulates in the bulk of the condensate, while the phase in the open gate region rounds up this bulk phase to properly match the circulation discreteness. Such overall behavior suggests the collective nature of the vortex oscillations.

The inclusion of dissipation, naturally, leads to the decay of persistent current oscillations over time. In the long run, the difference in angular momentum per particle decays to zero, trapping the vortex in the system's center [Fig.~\ref{fig: scheme}(c)]. This is an open-gate stationary equilibrium state, in which the angular momentum is distributed equally among the (equal radius) rings and persistent current oscillations are absent. Therefore, persistent current oscillations can be presented as excitations around this stationary state. In the case of small dissipation, oscillations gradually decay, as can be seen in Fig.~\ref{fig: scheme}(iv)-(v). (Note that while the oscillations in the angular momentum difference persist in such regime over the entire probed timescale (Fig.~\ref{fig: scheme}(v)), their amplitude after  $t\gtrsim 0.4\, \rm{s}$ becomes very small, thus also explaining the visible glitches of the winding number in Fig.~\ref{fig: scheme}(iv).) As anticipated, the lifetime of oscillations decreases rapidly with higher dissipation ($\gamma$), however for $\gamma>\gamma_{cr}\approx 0.015$ \cite{Bland_2022} we observe different behavior: the oscillations cease, and persistent current stays trapped in the initial ring, see Fig.~\ref{fig: scheme}(vi)-(vii). 

These ``perturbative" conclusions are also supported by our work of Ref.~\cite{chaika2024acceleration}, where we studied the influence of system acceleration, as an asymmetric parameter, on persistent current oscillations. We showed that the main role of acceleration was to introduce a stationary bias in oscillations. For the open-gate stationary state, the
phase accumulates faster in the less populated ring, so
the forward one, to properly satisfy the continuity equation. This leads to an angular momentum per particle bias in the complete protocol with oscillation dynamics: the vortex spends more time in the forward ring than in the backward one, while the overall period stays almost unchanged. Moreover, we showed that the component along the symmetry's main axis mostly defines the overall dynamics. This was expected, as the orthogonal acceleration component in the plane redistributes density equally within each ring. 

For some cases, we also included a closure stage in the overall protocol, where we closed the gate by linearly reducing the barrier potential [Fig.~\ref{fig: scheme}(i)]. This results in the vortex potentially localizing in a distinct ring. As shown, the process occurs smoothly, with the final vortex position consistent with the angular momentum immediately before the closure protocol. In the presence of simultaneous acceleration and sufficient dissipation, the vortex always settles in the forward ring—precisely where the phase bias occurs—regardless of its initial position, as illustrated in Fig.~\ref{fig: scheme}(viii)–(ix). These observations further support the idea that persistent current oscillations are ``carried" by system excitations.

\section{Normal modes analysis}

As mentioned earlier, persistent current oscillations can be viewed as disturbances to the open-gate system. Thus, analyzing elementary excitations can help explain the observed phenomena. Here, we employ the Bogoliubov-de Gennes (BdG) formalism \cite{pitaevskii2016bose,achilleos2012dark}, to describe collective excitations. As we will demonstrate later, this formalism adequately explains the observed oscillations. Therefore, we can conclude {\it a posteriori} that the opening and closing parts of our standardized protocol cause a small disturbance. The condensate wavefunction takes the form of
\begin{equation*}
\psi(\boldsymbol{r},t)=\psi_0(\boldsymbol{r})+\delta \psi(\boldsymbol{r},t)
\end{equation*}
where $\psi_0(\boldsymbol{r})$ is the stationary state with corresponding chemical potential $\mu_{\rm{2D}}$, and $\delta \psi(\boldsymbol{r},t)$ is a small perturbation of the form
\begin{equation*}
    \delta \psi(\boldsymbol{r},t) =\sum_k c_k \left[ u_k(\boldsymbol{r}) e^{-i \omega_k t} +{v}_k^{*}(\boldsymbol{r}) e^{i {\omega}_k^{*} t}\right].
\end{equation*}

Inserting this into Eq.~\eqref{eq:dGPE2D}, and taking into account first-order perturbations, one obtains the following system of Bogoliubov-de Gennes equations
\begin{equation} \label{eq: BdG system}
    \begin{aligned} 
       \hbar \omega_k (1+i \gamma) u_k = \left[\hat{H}-\mu_{\rm{2D}}+2g|\psi_0|^2\right] u_k + g \psi_0^2 v_k,\\
  -\hbar \omega_k (1-i \gamma) v_k = \left[\hat{H}-\mu_{\rm{2D}}+2g|\psi_0|^2\right] v_k + g {\psi_0^{*}}^2 u_k,
\end{aligned}
\end{equation}
where
\begin{equation*}
    \hat{H}=-\frac{\hbar^2}{2M} \nabla^2 +V(\boldsymbol{r}),
\end{equation*}
and $V(\boldsymbol{r})=V_{ext}(\boldsymbol{r})+M \boldsymbol{a} \cdot \boldsymbol{r}$ is the total potential, including the acceleration correction, with the usual normalization condition
\begin{equation}
    \int \,(|u_k|^2-|v_k|^2)\text{d}x\text{d}y=1.
\end{equation}
This linear system for the eigenmodes \eqref{eq: BdG system} can be solved numerically (see, for example, Ref. \cite{gao2020numerical}) for different system parameters, such as acceleration, dissipation, and the total winding number. Such results are shown in this section, and are compared to results of the full GPE real-time simulations. 

Typically, only the lowest energy excitations are significantly excited and play a crucial role in perturbative dynamics. As it was found {\it a posteriori}, due to our one-dimensional (1D)-like condensate geometry, the lowest modes have a sound-like behavior, which should be captured by a hydrodynamic approach, widely used in the literature \cite{Capuzzi, Zaremba_1998,abad2014persistent, nikuni1998hydrodynamic, stringari1996collective}. In the following subsections, we consider an effective acoustic model that mimics the original open system and gives a qualitative analytical explanation of the eigenmode behavior.

\subsection{Effective 1D acoustic model}
Here, we analyze the normal modes of the following simplified model in the Thomas-Fermi approach by approximating the system as a strip-like configuration having periodic boundaries along a new axis ``$z$", where $z \in [0, L)$, while being confined under the harmonic trap $V={M \omega_r^2 r_{\perp}^2}/{2}$ in a perpendicular $r_{\perp}$ direction. Also, we define that the initial system's left and right ring correspond to the left $z \in [0, L/2)$, and right part $z \in [L/2, L)$ of this strip. Essentially, we have reduced our full system to a ``straightened version" of the original system, a thin annulus, broadly studied elsewhere \cite{cozzini2006vortex, kumar2016minimally, Polo2018}.
Here, we extend the hydrodynamic analysis by including dissipation and accounting for our overall setup of excitation measurement. The system length is estimated as the length of the peak density of the initial one. It implicitly depends on the chemical potential and the barrier amplitude, and can be calculated in the Thomas-Fermi approximation. For our parameters, this dependence is negligible, of order $1\%$, and we take a length equal to $L=4 \pi \times0.88 R$ to account for the overlap region. 

The GPE \eqref{eq:dGPE2D} can be recast as the following system of hydrodynamic equations, through the Madelung transformation [$\psi=\sqrt{\rho}e^{i \Phi}$, also note that in a stationary state, the $\partial_t \Phi=0$  according to \eqref{eq:dGPE2D}]
\begin{equation*} 
\begin{aligned}
        &\frac{\partial \rho}{\partial t} + \nabla \cdot (\rho \boldsymbol{v})= 2\rho \gamma \frac{\partial{\Phi}}{\partial t}, \\
        &g \rho + V+\frac{M}{2}v^2-\frac{\hbar^2}{2M} \frac{\nabla^2 \sqrt{\rho}}{\sqrt{\rho}}  = \hbar \left[ \mu_{\rm{2D}} -\frac{\partial{{\Phi}}}{\partial t} - \gamma \frac{1}{ 2 \rho} \frac{\partial \rho}{\partial t} \right]\,.
\end{aligned}
\end{equation*}
Here $\rho$ is the density and $\Phi$ the phase, while the equations represent continuity and momentum conservation laws, respectively. Hereafter, we consider only small values of the effective dissipation $\gamma \ll 1$, which is physically relevant for ultracold atomic gases ~\cite{weiler2008spontaneous,rooney_2013,Liu18Dynamical,Ota18Collisionless,szigeti-ring}. Additionally, within this consideration, we assume the smallness of the persistent current velocities compared to the speed of sound and neglect the effect of acceleration, which alters the ground state and excitation spectrum.
Therefore under such assumptions, the terms $\frac{\hbar^2}{2M} \frac{\nabla^2 \sqrt{\rho}}{\sqrt{\rho}}$ and $\gamma \frac{1}{ 2 \rho} \frac{\partial \rho}{\partial t}$ can be dropped, within the acoustic regime. 

For the sake of simplicity, hereafter we proceed to work in harmonic units, $t\rightarrow t\omega_r$, $z \rightarrow z/l_r$, $\mu \rightarrow \mu_{\rm{2D}}/\hbar \omega_r$, where $l_r=\sqrt{\hbar/(M \omega_r)}$, therefore we have the following hydrodynamic equations in the Thomas-Fermi approximation
\begin{equation} \label{eq: Madelung 1}
\begin{aligned}
            \frac{\partial \rho}{\partial t} + \nabla \cdot (\rho \boldsymbol{v})= 2\rho \gamma \frac{\partial{\Phi}}{\partial t}, \\
        g \rho +\frac{1}{2}r_{\perp}^2+\frac{1}{2}v^2= \mu -\frac{\partial{{\Phi}}}{\partial t}.
\end{aligned}
\end{equation}
The stationary state profile corresponds to an inverse parabola in the direction of $r_{\perp}$, and the chemical potential can be estimated as $\mu= \left(\frac{3 g N}{4 \sqrt{2} L} \right)^{2/3} $. We also assume the nonzero persistent current velocity $\boldsymbol{v}= v_0 \boldsymbol{e_z}$, corresponding to a present vortex with winding number $m$, as $v_0=m \times 2\pi/L $.

By linearizing \eqref{eq: Madelung 1} around the given stationary state, and assuming the following form of perturbations
\begin{align*}
    &\delta \rho(\boldsymbol{r},t)=\delta \rho (r_{\perp}) e^{i (q z-\omega t)}, \\
    &\delta \Phi(\boldsymbol{r},t)=\delta \Phi (r_{\perp}) e^{i (q z-\omega t)}
\end{align*}
we get the following system for $\delta \rho (r_{\perp})$ and $\delta \Phi (r_{\perp})$
\begin{equation} \label{eq: Madelung 2}
 \begin{aligned} 
  &  i\left(v_0 q -\omega\right) \delta \Phi (r_{\perp}) =-g \delta \rho (r_{\perp}), \\
  & i\left(v_0 q -\omega\right) \delta \rho (r_{\perp})=q^2 \rho_0(r_{\perp}) \delta \Phi (r_{\perp}) \\ &-2 i \rho_0(r_{\perp}) \gamma \omega \delta \Phi (r_{\perp}) -\nabla_{\perp} \left(\rho_0(r_{\perp}) \nabla_{\perp}\delta \Phi (r_{\perp}) \right ).
\end{aligned}   
\end{equation}
In the case of $q\rightarrow0$, the lowest-energy solution has a constant radial profile along $r_{\perp}$ axis, i.e. $\delta \rho(r_{\perp})$ and $\delta \Phi(r_{\perp})$ constant. Therefore, for the case of small wavenumber $q \ll 1$, we can ignore its influence on radial profile structure, and integrate equations \eqref{eq: Madelung 2} within the perpendicular Thomas-Fermi profile, similar to \cite{Zaremba_1998, Capuzzi,abad2014persistent}. Hence, the system \eqref{eq: Madelung 2} simplifies to
\begin{equation} \label{eq: wave equation}
   \begin{aligned} 
  &  i\left(v_0 q -\omega\right) \delta \Phi =-g \delta \rho,  \\
  & i\left(v_0 q -\omega\right) \delta \rho =\langle \rho_0 \rangle \left(q^2  - 2 i \gamma \omega \right) \delta \Phi\,, 
\end{aligned} 
\end{equation}
whose solution corresponds to the following dispersion relation
\begin{equation} \label{eq: dispersion relation}
    (\omega-v_0 q)^2=g\langle \rho_0 \rangle \left(q^2  - 2 i \gamma \omega \right).
\end{equation}
Here $\langle \rho_0 \rangle$ is the average density in the radial  Thomas-Fermi region. The factor $g \langle \rho_0 \rangle=2/3 \mu$ corresponds to the square of the speed of sound, in the conservative case; the latter is equal to $c=\sqrt{2/3} \,c_B$, where $c_B$ is the Bogoliubov speed of sound for the homogeneous case with the same peak density. If we initially consider a 3D cylinder configuration, instead of a quasi-2D geometry, we would get the factor $c_{\rm{3D}}={1/\sqrt{2}} \,c_B$ \cite{Zaremba_1998, abad2014persistent}. The wavenumber is $q=2\pi n/L$, where $n$ is an integer, due to periodic boundary conditions. Therefore, from \eqref{eq: wave equation} and \eqref{eq: dispersion relation}, we establish a complete set of eigenvectors and corresponding eigenmodes.

\begin{figure*}[!ht] 
   \centering   
   \includegraphics[width=2\columnwidth]{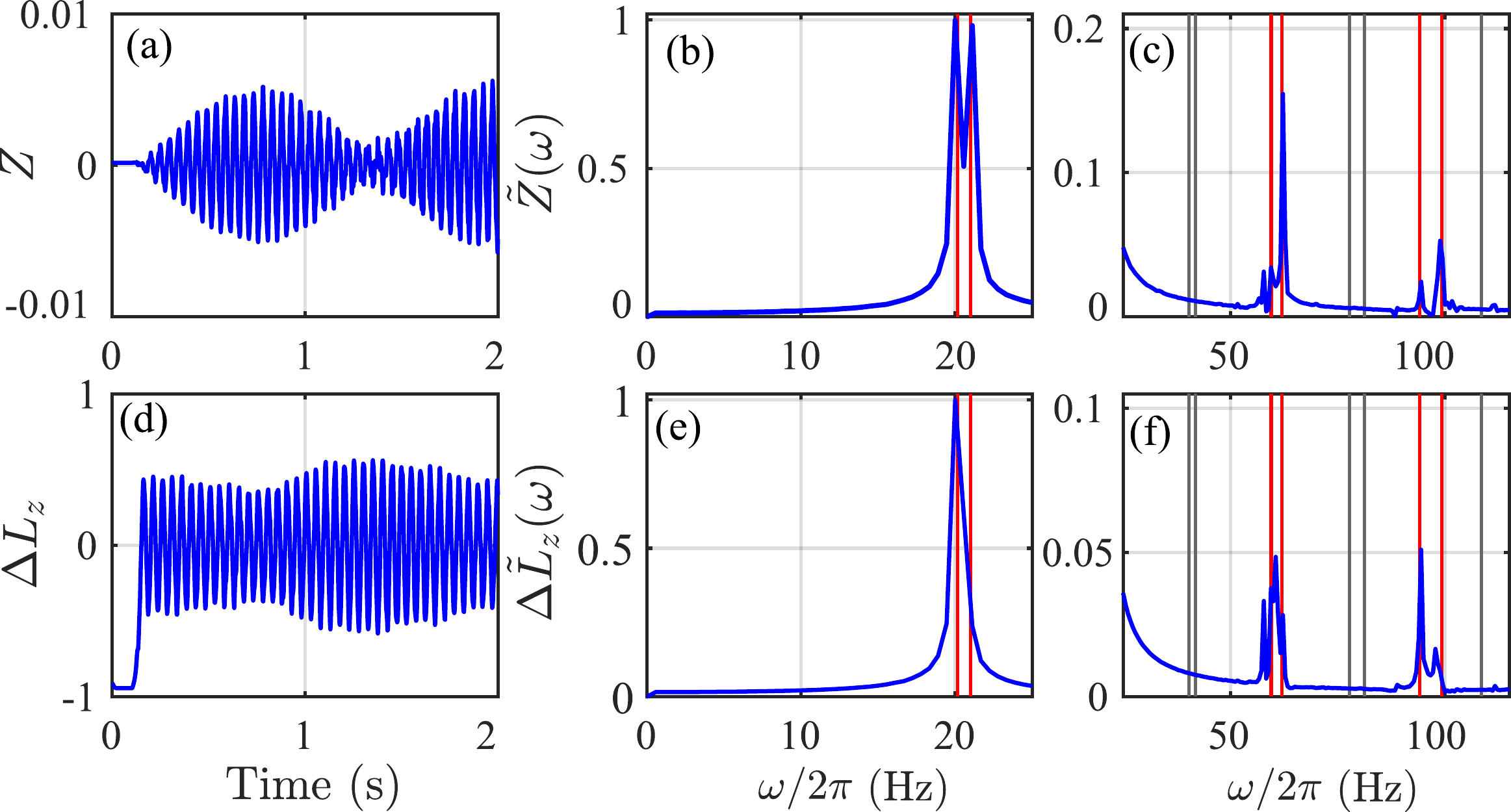}%
\caption{Analytical and numerical predictions of beating effects in vortex oscillations. (a) Population imbalance between rings and (d) angular momentum per particle difference dynamics for the opening gate protocol at $V_0=1.2 \mu$, obtained in numerical simulations of conservative GPE \eqref{eq:dGPE2D}. (b) and (c) corresponding Fourier spectra, normalized to the maximum value, of population imbalance (a), at different frequency ranges. Note how much smaller the scale of (c) is relative to (b). 
(e) and (f) similar normalized Fourier spectra of angular momentum difference (d). Vertical lines denote the corresponding elementary excitation \eqref{eq: BdG system} of the open-gate stationary solution. Red lines denote visible, while grey lines denote inactive modes.}
\label{fig: imbalance dynamics single vortex}
\end{figure*}

However, depending on what we are measuring and what
the symmetries of the system and the measurement operator are, only some modes can be visible. In practice, we have measured particle and phase imbalance, and as we discussed for the latter, only the phase imbalance in the bulk is practically relevant. The phase imbalance ($\Delta \Phi=\Phi_{\rm{Left}}-\Phi_{\rm{Right}}$), for our effective acoustic model, can be presented as
\begin{equation*}
\begin{aligned}
           &\Phi_i = \int_i v \text{d} z = \int_i (v_0+\delta v) \text{d} z \rightarrow \\ &\Delta \Phi= \left(\int_{\rm{Left}} v_0 \text{d} z -\int_{\rm{Right}} v_0 \text{d} z\right)+2 \int_{\rm{Left}} \delta v \text{d} z \\ &= \Delta \Phi_0+2 \delta \Phi|_\text{edges}. 
\end{aligned}
\end{equation*}
Here, somehow schematically, we expressed the $\Phi_i$ in each ring by explicitly highlighting the stationary and perturbative phase distribution. Also, we used the fact that the total perturbation equals zero. The same can be done for the number imbalance to give
\begin{equation*}
\begin{aligned}
           &N_i = \int_i \rho \text{d} z = \int_i (\rho_0+\delta \rho) \text{d} z \rightarrow \\ &\Delta N= \left(\int_{\rm{Left}} \rho_0 \text{d} z -\int_{\rm{Right}} \rho_0 \text{d} z\right)+2 \int_{\rm{Left}} \delta \rho \text{d} z \\ &= \Delta N_0 +2 \int_{\rm{Left}} \delta \rho \text{d} z.
\end{aligned}
\end{equation*}

The $\Delta \Phi_0$ and $\Delta N_0$ present a stationary bias, which is present in the case of some asymmetric factor, like acceleration, that we ignore for now, so both of them are zero. So, the present imbalance is purely perturbative, and the eigenmodes, which are symmetric to the main axis, stay inactive. Therefore, all even modes in our acoustic model stay inactive in the imbalance dynamics. For the conservative regime, using \eqref{eq: wave equation} and \eqref{eq: dispersion relation}, the phase and density imbalance can be written as 
\begin{equation}  \label{eq: imbalance}
\begin{aligned}
        \Delta N=\sum_{q>0, \, \text{odd}} \left[A_q e^{-iq(c+v_0)t}+A_q^{\ast} e^{iq(c+v_0)t}\right] \\ 
        +\left[B_q e^{iq(c-v_0)t}+B_q^{\ast} e^{-iq(c-v_0)t}\right], \\
        \Delta \Phi= \frac{g}{c}\sum_{q>0, \, \text{odd}} \left[A_q e^{-iq(c+v_0)t}+A_q^{\ast} e^{iq(c+v_0)t}\right] \\ 
        -\left[B_q e^{iq(c-v_0)t}+B_q^{\ast} e^{-iq(c-v_0)t}\right]\,.
\end{aligned}
\end{equation}
Here, $A_q$ and $B_q$ are complex constants determined by the initial conditions, i.e. the initiation protocol. Physically, their amplitudes and phases correspond to those of waves propagating around the system in opposite directions. In the presence of a persistent current (vortex), the frequencies of these waves are shifted, $\omega_{\pm} = q(c \pm v_0)$, analogous to the standard Sagnac interference setup. For our parameters, the velocity of the persistent current is much smaller than the speed of sound--for a single vortex, $v_0 \approx 0.014 c$--and therefore contributes only a small correction to the eigenfrequencies. Nonetheless, the role of the vortex is crucial, as it breaks the symmetry of the system. If one were to consider the same setup without vorticity, both the phase and density imbalances would vanish due to the symmetry of the protocol. In such a case, only even (and thus inactive) waves could be generated, in agreement with our observations.

\subsection{Comparison of GPE simulations with BdG model in the conservative regime}
\begin{figure}
   \centering   
   \includegraphics[width=\columnwidth]{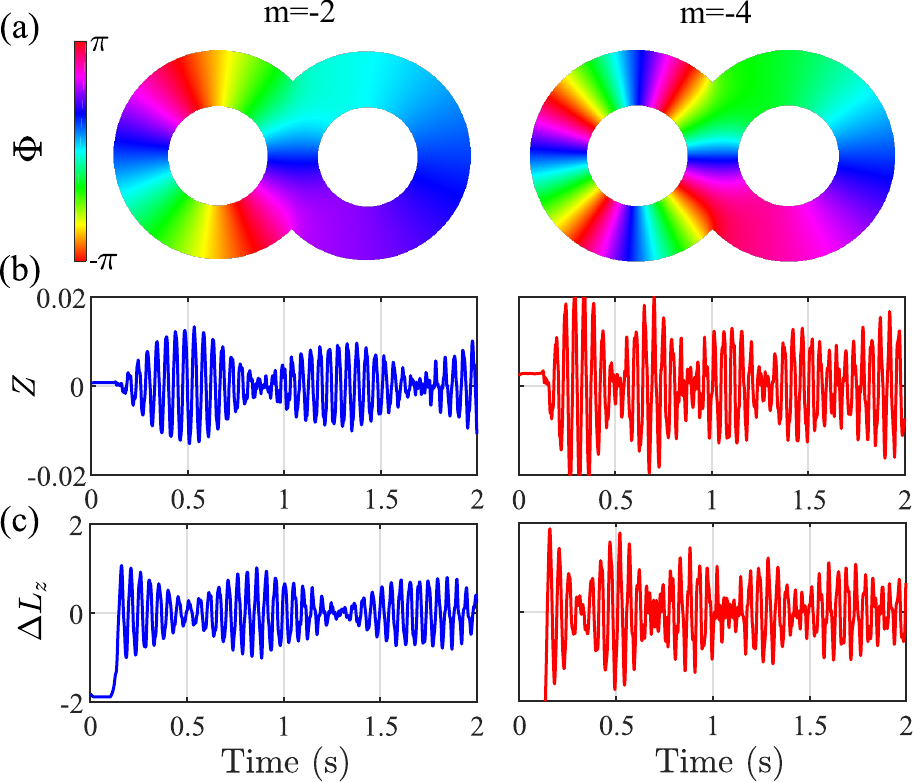}%
\caption{Oscillations and beating effects of multiply charged persistent currents. (a) initial phase, (b) population imbalance, and (c) angular momentum per particle difference dynamics for the opening gate protocol for two (left column) and four (right column) vortices in the system.}
\label{fig: few vortex beatings}
\end{figure}

Here, we verify our normal mode analysis for the simplest, but defining, case of the absent acceleration and dissipation [Fig.~\ref{fig: scheme}(b)]. In Fig.~\ref{fig: imbalance dynamics single vortex}(a) and Fig.~\ref{fig: imbalance dynamics single vortex}(d), we present long-time dynamics of particle imbalance between rings and angular momentum difference, for the open-gate protocol, obtained by numerical simulations of \eqref{eq:dGPE2D}, with a single vortex, initially placed in the left ring. The Fourier spectrum [Fig.~\ref{fig: imbalance dynamics single vortex}(b)] shows that the spectrum is dominated by two close, almost merged peaks, around $20\, \rm{Hz}$, with minor influence from odd overtones around $60\, \rm{Hz}$, and $ 100\, \rm{Hz}$ correspondingly. Additionally, we observe that qualitatively, the spectra for angular momentum difference and particle imbalance are the same, but the relative amplitudes differ. Alongside real-time dynamics [Fig.~\ref{fig: imbalance dynamics single vortex}(b), Fig.~\ref{fig: imbalance dynamics single vortex}(e)], we also show the spectrum of the BdG system [Eq.~\eqref{eq: BdG system}], for the open-gate stationary solution, which we obtain through additional real-time evolution with small dissipation. This low-energy spectrum has phonon-like behavior, in agreement with the effective acoustic model. First of all, one can discern different pairs of BdG modes, where each one is characterized by the same number of nodes along the condensate peak density (has the same $q$ relative to the acoustic model \eqref{eq: imbalance}). Such pairs have similar eigenfrequencies [see five consecutive pairs in Fig.~\ref{fig: imbalance dynamics single vortex} (b) and Fig.~\ref{fig: imbalance dynamics single vortex}(c)]. The lowest and dominating pair has frequencies $\omega_{\pm} = 2\pi \times (20.57 \pm 0.42) \, \rm{Hz}$, while the prediction of the acoustic model gives estimate $\omega_{\pm} = q(c \pm v_0)\approx2\pi \times (19.47 \pm 0.27) \, \rm{Hz}$. We deliberately present these frequencies in such a form because such modes typically have similar amplitudes, thus generating a characteristic beating pattern, visible in Fig.~\ref{fig: imbalance dynamics single vortex} and Fig.~\ref{fig: few vortex beatings}. Therefore, the average frequency characterizes fast dynamics, while frequency splitting characterizes slow envelope dynamics.

As shown in Fig.~\ref{fig: imbalance dynamics single vortex}, even modes, marked by grey lines, indeed stay inactive in the overall imbalance dynamics, as we expected within the effective model approach. To properly characterize the symmetries of the BdG modes, we used the same approach as in Ref.~\cite{momme2019collective}, in which the authors introduced the population imbalance operator
\begin{equation*}
    Z=\frac{1}{N}\langle \Psi | \hat{Z}|\Psi \rangle, \quad \quad \hat{Z}(\boldsymbol{r})= 
    \begin{cases} 
\quad 1, & \text{Right  ring: } (x > 0), \\
\: -1, & \text{Left  ring: } (x < 0).
\end{cases}
\end{equation*}
For the present case of a BdG perturbation, this turns into the following excitation coefficient
\begin{equation*}
    Z_k=\bigg| \int \hat{Z}\left(\psi_0 v_k +{\psi_0^{*}} u_k\right) \text{d}V\bigg|. 
\end{equation*}
We have directly verified that such excitation coefficients $Z_k$ are significantly smaller for even modes, relative to odd modes, by two orders of magnitude, so we characterize them as inactive in terms of imbalance measurement.

For the given example of the single vortex, the frequency split obtained by BdG properly agrees with the corresponding real-time imbalance dynamics, while the angular momentum difference dynamics does not have a clear beating pattern [Fig.~\ref{fig: imbalance dynamics single vortex}(d)]. However, for examples with more vortices in the system, as in Fig.~\ref{fig: few vortex beatings}, beatings are more pronounced for both measured values, but the overall patterns still modify with time. Such behavior could be attributed to the higher-order interaction between modes, which was ignored in our analysis.

\begin{figure}
   \centering   
   \includegraphics[width=\columnwidth]{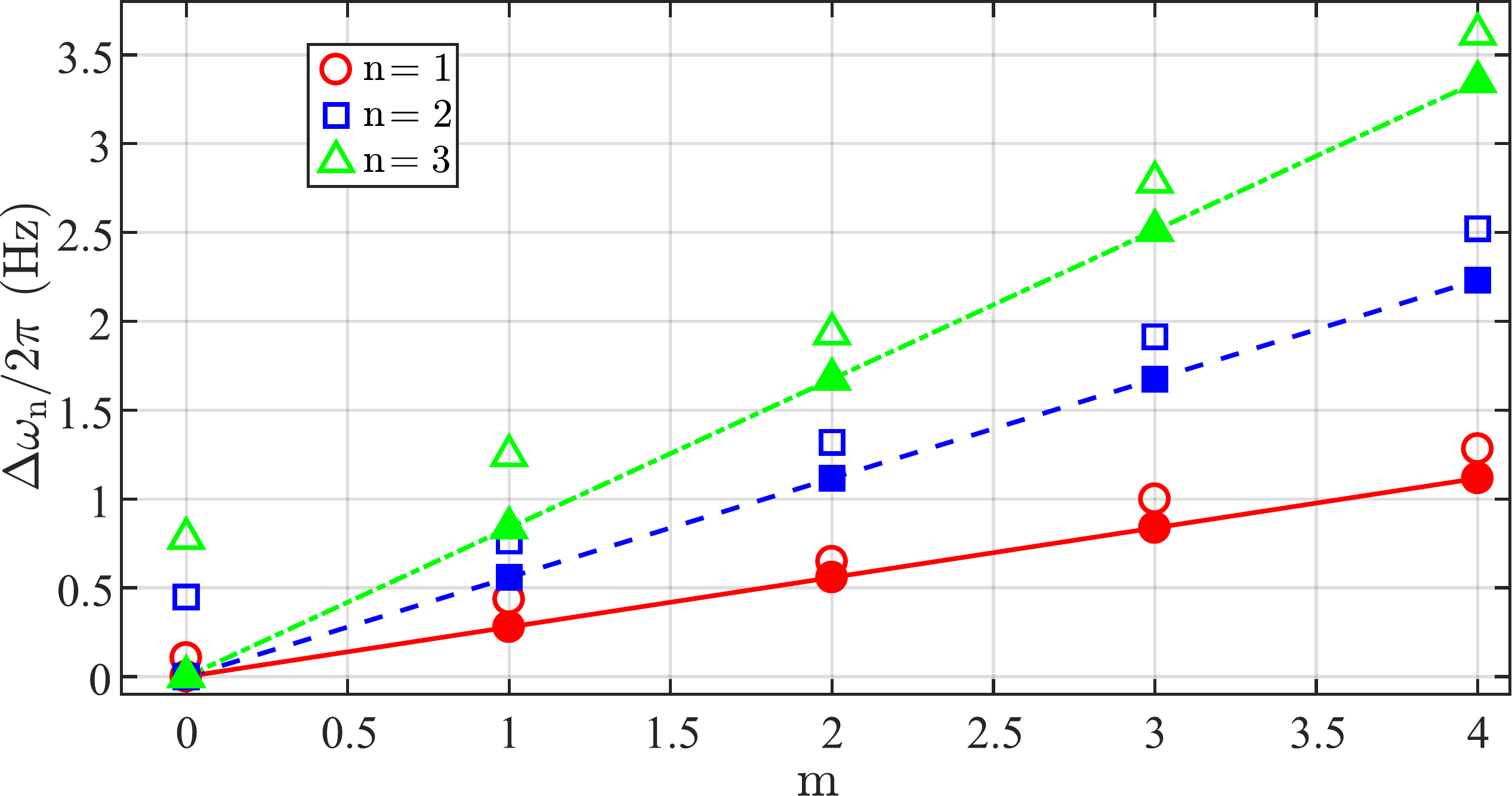}%
\caption{The frequency split (envelope frequency) dependence on the present vorticity in the system ($m$), for different acoustic waves $\rm{n}$. The points present the solution of \eqref{eq: BdG system}; while the lines show the acoustic model approach, whenever filled points correspond to quantized vorticity values.}
\label{fig: beating frequency}
\end{figure}

As we can see, the effective acoustic model provides a good estimate of the average frequency, within $5\%$, but a looser estimate of the frequency split, within a factor of two. In Fig.~\ref{fig: beating frequency}, we show how the frequency split depends on phonon wavenumber $n$, and the number of vortices in the system $m$. As can be seen, the frequency split is present even when there is no persistent current in the system. 
In the case of absent vorticity, the BdG modes have a standing wave structure. Due to our geometry, which has the major and minor symmetry axes (along the $x$- or $y$-axis, respectively, see Fig.~\ref{fig: scheme}), the nodes of the eigenmodes are located on one of these symmetry axes, predictably having different structures and frequencies. This ``geometric" contribution remains when the persistent current is introduced, affecting the overall frequency split, while the BdG modes turn into a superposition of counter-propagating waves with unequal amplitudes. Nevertheless, for higher vorticity, this ``geometric" effect is less significant, and the frequency split value aligns better with acoustic model estimates.

\begin{figure}
   \centering   
   \includegraphics[width=\columnwidth]{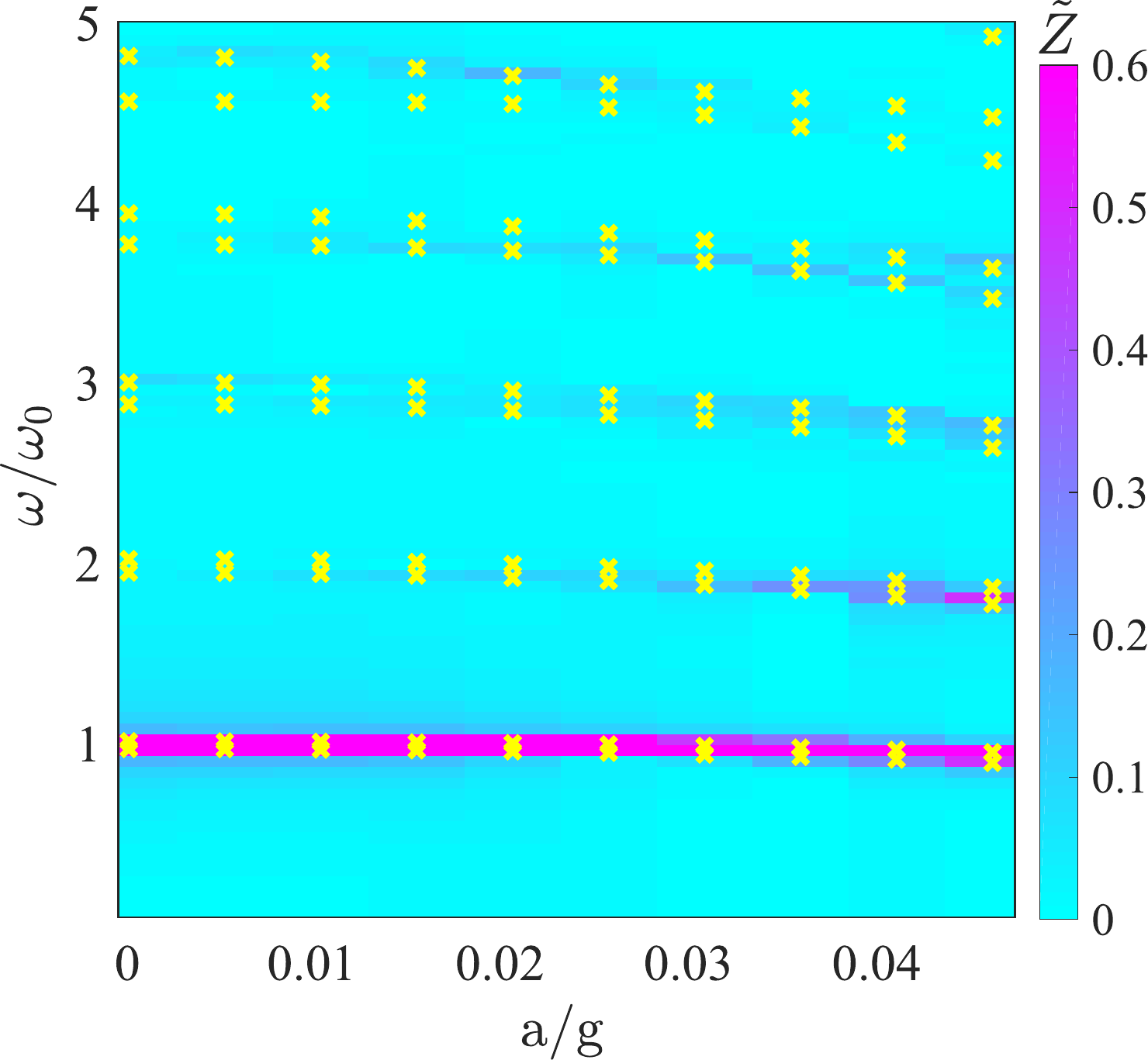}%
\caption{The normalized Fourier spectrum of the GPE \eqref{eq:dGPE2D} without dissipation ($\gamma=0$). Shown is the particle imbalance dynamics, as a function of the present acceleration along the main axis in the system. Yellow crosses present numerical results of respective BdG eigenfrequencies \eqref{eq: BdG system} of the open-gate stationary solution, at given acceleration. More pinkish regions correspond to more active modes. The vertical (frequency) axis is normalized to $\omega_0=2\pi \times 20.57 \, \rm{Hz}$, which is the average frequency of the lowest phonon modes.}
\label{fig: acceleration}
\end{figure}

Therefore, as we can see for the conservative case, the normal-mode analysis of the open-gate stationary state accurately describes the characteristic frequencies observed in full real-time dynamics. The relative amplitudes of the modes lie beyond such analysis, as they depend on the barrier-opening protocol; however, our numerous simulations with different parameters show that, under the standard opening protocol, the lowest pair of modes is predominantly excited. Additionally, we note that mode mixing naturally presented within our non-linear system does not disrupt the qualitative low-energy acoustic picture. We have verified that spectral peaks extracted from the full nonlinear time-dependent GPE simulations track the BdG eigenfrequencies [e.g. Fig.~\ref{fig: imbalance dynamics single vortex}(a) and Fig.~\ref{fig: imbalance dynamics single vortex}(b-c)] and decay rates of the stationary open-gate state over a broad range of parameters, including variations of acceleration and dissipation, which we cover in the following sections [Fig.~\ref{fig: acceleration}]. This quantitative tracking shows that, while nonlinearity can
redistribute spectral weight among modes, it does not alter the identification of the dominant low-energy acoustic
modes responsible for circulation exchange and damping, nor the predicted dependence on system parameters. Moreover, the simplified analytical acoustic model also captures these features quite well; its success arises from the effectively 1D nature of the system, as highlighted in \cite{kumar2016minimally}. One way to improve this model is to include the spatially dependent curvature of the waveguide, which we have neglected but which has been addressed in Refs.~\cite{sandin2017dimensional, campo2014bent, salasnich2022bose, schwartz2006one}. Incorporating such effects would allow for a more accurate description of frequency splitting, providing another avenue for studying beating effects \cite{cozzini2006vortex, marti2015collective, kumar2016minimally, bayocboc2023frequency}.

\subsection{Influence of acceleration on eigenmodes} 
The main role of acceleration is to introduce a static bias in the phase and particle imbalance. Here, we examine more subtle effects of acceleration on the normal modes. In Fig.~\ref{fig: acceleration}, we present the Fourier spectrum of the particle-imbalance dynamics, obtained numerically from GPE simulations (as in Fig.~\ref{fig: imbalance dynamics single vortex}(b)) for different values of acceleration. To highlight the role of excitations, the stationary shift of the particle imbalance was removed prior to the Fourier transform. Overall, the spectrum again agrees with the corresponding BdG eigenvalues of the open-gate protocol (yellow crosses in Fig.~\ref{fig: acceleration}). All eigenfrequencies exhibit only a weak dependence on acceleration, decreasing monotonically in a roughly quadratic fashion, and differing by no more than about $10\%$ even near the critical acceleration. Interestingly, the visibility of modes shows a strong dependence on acceleration: even modes become increasingly prominent at higher acceleration, while the lowest pair of acoustic modes ($q=1$) remains dominant across the explored range. Such behavior lies beyond our effective acoustic model, but can be qualitatively reproduced by including an additional acceleration potential $V \sim \cos(2\pi x/L)$ within the hydrodynamic model \cite{cozzini2006vortex}, which can be treated numerically or perturbatively. We do not pursue these details further here, as their impact is relatively minor.

\subsection{Oscillation decay within phenomenological dissipation model}

While the results of normal-mode analysis are relatively familiar in the conservative case--similar to single-ring setups \cite{cozzini2006vortex, kumar2016minimally}, here we focus on the influence of dissipation in our double-ring geometry. In the literature, phenomenological dissipation is often introduced in a simplified form to describe dissipative effects in Bose–Einstein condensate dynamics \cite{choi1998phenomenological, tsubota2002vortex, Bradley2012}.
The effect of $\gamma$ on the excitation spectrum has been studied less extensively, and is most commonly discussed in the context of soliton and vortex anti-damping behavior \cite{achilleos2012dark,yan2014exploring,middelkamp2010stability,carretero2016vortex,cockburn2011fluctuating,cockburn2010matter}. In contrast, Ref.~\cite{wouters2012energy} obtained essentially the same phonon dispersion relation \eqref{eq: dispersion relation} within the BdG framework, but explored the role of dissipation more thoroughly in polariton condensates. Motivated by this, we investigate how $\gamma$ modifies the phonon spectrum in our system.

\begin{figure}
   \centering   
   \includegraphics[width=\columnwidth]{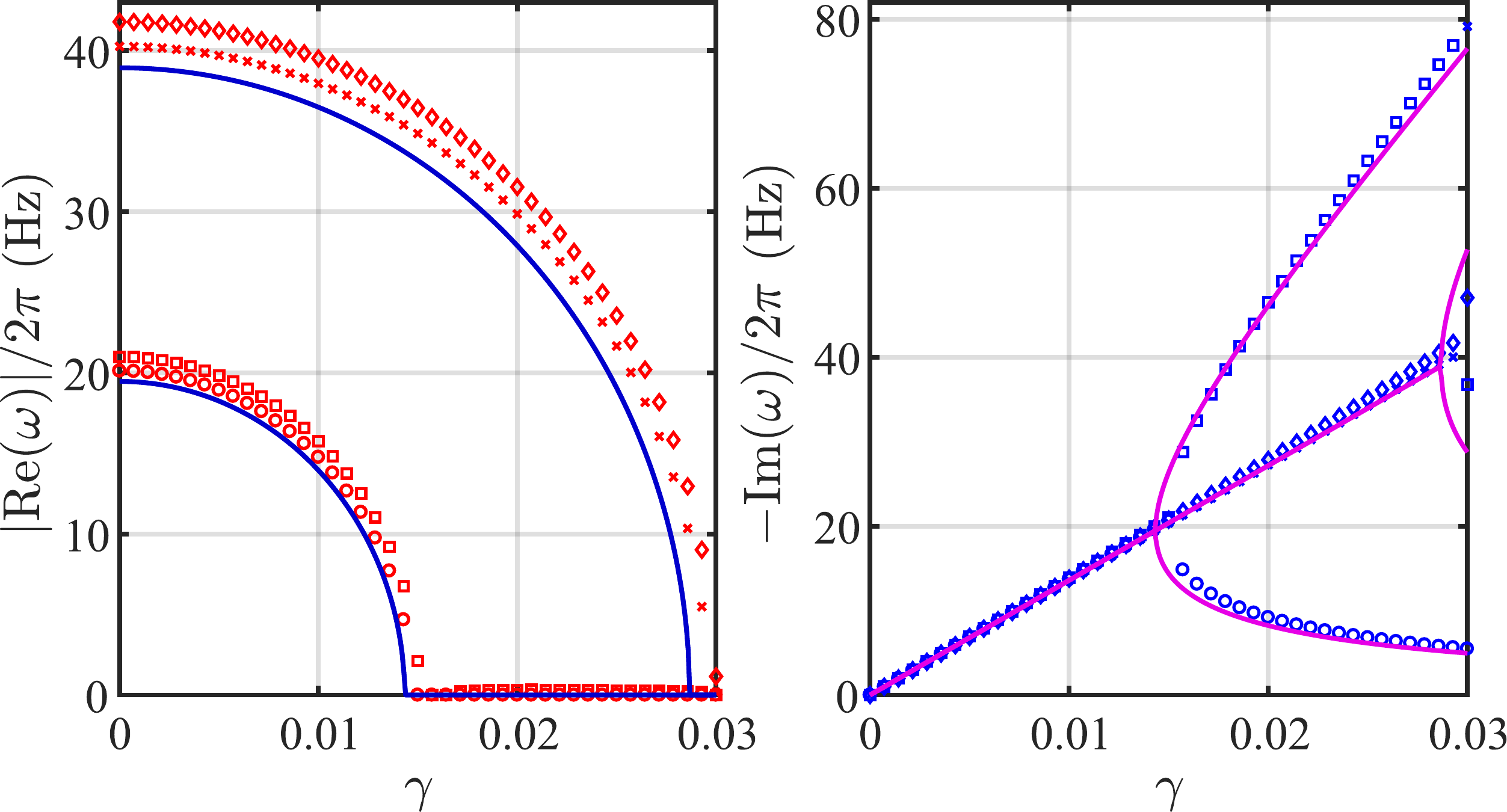}%
\caption{Excitation spectrum of real (left) and imaginary part (right) of the lowest two acoustic modes for the open-gate stationary solution, as a function of dissipation parameter $\gamma$. The points present the numerical solution of \eqref{eq: BdG system}, while the lines show the effective acoustic model approach \eqref{eq: dissipative dispersion 2}. Squares and circles denote the $n=1$ mode, while crosses and diamonds denote the $n=2$ mode.}
\label{fig: dissipative eigenmodes}
\end{figure} 

We start our analysis with the effective acoustic model prediction \eqref{eq: imbalance}. As mentioned earlier, in the case of a single vortex the persistent current can be neglected, since $v_0 \approx 0.014c$. The dispersion relation \eqref{eq: dispersion relation} therefore simplifies to
\begin{equation} \label{eq: dissipative dispersion 2}
\omega = - i \gamma g \langle \rho_0 \rangle \pm \sqrt{g \langle \rho_0 \rangle q^2 - (g \langle \rho_0 \rangle \gamma)^2}.
\end{equation}
For small $\gamma$, all sound waves acquire a decay rate $\Gamma_q = \gamma g \langle \rho_0 \rangle$, which is predictably proportional to $\gamma$. However, as $\gamma$ increases, the system enters an overdamped regime: the frequencies bifurcate into two purely damped modes at the critical value
\begin{equation}
\gamma_{cr} = n \frac{2 \pi}{L \sqrt{g \langle \rho_0 \rangle}} \approx n \times 0.014.
\end{equation}
For the lowest mode ($n=1$), this numerical value coincides with the ratio of the persistent current velocity to the speed of sound ($v_0/c \approx 0.014$), and is remarkably close to the overdamped regime observed in \cite{Bland_2022}, occurring for $\gamma > \gamma_{cr} \approx 0.015$. Increasing $\gamma$ further modifies the decay rate as
\begin{equation*}
\Gamma_q = g \langle \rho_0 \rangle \left(\gamma \pm \sqrt{\gamma^2 - \gamma_{cr}^2}\right).
\end{equation*}

When $\gamma \gg \gamma_{cr}$, the decay rates become $\Gamma_{q1} \approx 2 g \langle \rho_0 \rangle \gamma$ and $\Gamma_{q2} \approx g \langle \rho_0 \rangle {\gamma_{cr}^2}/{(2 \gamma)}$. Thus, instead of two counter-propagating waves with wave number $q$ (as in the low-dissipation regime), the system supports two purely decaying--or “spatially frozen’’--modes. One of these modes decays rapidly, while the other decays more slowly and becomes dominant over all remaining modes. This is in agreement with observed overdamped dynamics in \cite{Bland_2022}, as well as with Fig.~\ref{fig: scheme}(vi) and Fig.~\ref{fig: scheme}(vii), where $\gamma=0.02$ was used. In Refs.~\cite{achilleos2012dark,cockburn2010matter}, a similar, but anti-damping, bifurcation is observed for soliton dynamics. Our acoustic model results with included dissipation also accurately match the direct numerical BdG spectrum, presented in Fig.~\ref{fig: dissipative eigenmodes}, especially in describing the decay rate behavior. For small values, all phonon pairs have similar decay rates, while at $\gamma \approx 0.015$ and $\gamma \approx 0.030$, the first and second pair bifurcate into an overdamped regime. We have also verified that the given decay rate complies with real-time simulations. We would note that the inclusion of persistent current presence modifies the spectrum, including decay rates, although for the studied case of a single vortex, it is indeed insignificant [Fig.~\ref{fig: dissipative eigenmodes}]. In Appendix A, we address this interplay more closely.

\section{Controlled persistent current transfer and experimental relevance}
\subsection{Resonant vortex transfer}

\begin{figure}
   \centering   
   \includegraphics[width=\columnwidth]{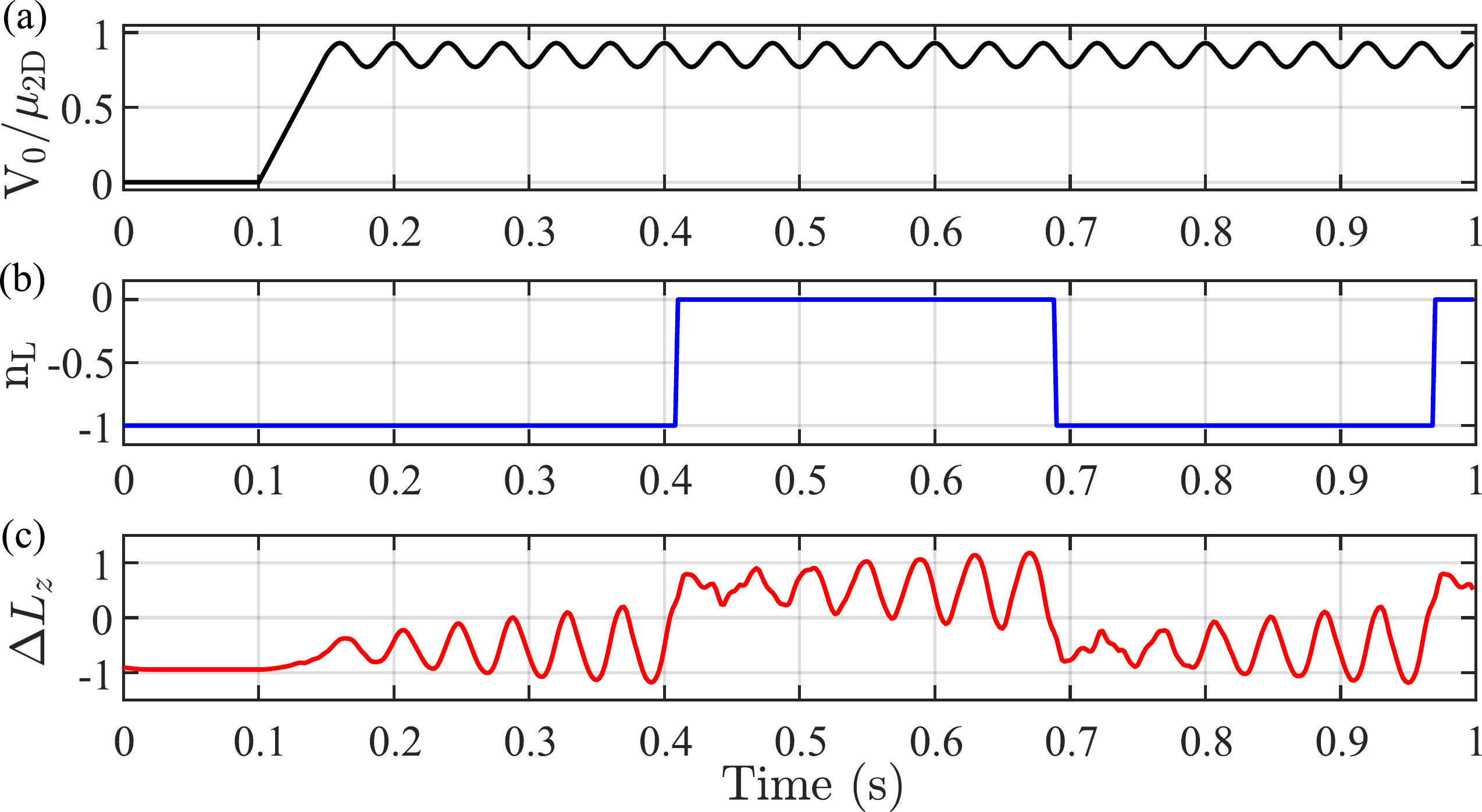}%
\caption{Example of resonant current transfer protocol, with dissipation $\gamma=0.001$, similar to Fig.\ref{fig: scheme}. Dynamics of (a) modulating barrier amplitude, (b) winding number at the left ring, and (c) angular momentum per particle difference. }
\label{fig: resonant transfer}
\end{figure}

The phononic nature of the persistent current oscillations revealed here provides a simple mechanism for enabling controlled phase slips in ring condensates: periodic modulation of the barrier potential at resonance with the phonon frequency, similar to the protocol employed in Refs.~\cite{wang2015resonant, kuriatnikov2020phase}, can allow vortex transport, even in the case of density-separated rings. In Fig.~\ref{fig: resonant transfer}, we demonstrate a successful example of a resonant protocol, where we periodically modulate the barrier amplitude \eqref{eq: trap barrier}, $V_0(t)$, during the open-gate stage of the protocol for the dissipative GPE system \eqref{eq:dGPE2D}. The maximum barrier amplitude $V_0 = 0.93 \mu_\text{2D}$ is well below the chemical potential, which, for the “standard’’ protocols with unmodulated barrier amplitude (as in Fig.~\ref{fig: scheme}), prohibits any vortex transitions due to the distinct two-ring topology. The resonant frequency here is $25 \, \rm{Hz}$, slightly different from the $\approx 21 \, \rm{Hz}$ of the open-gate protocol, as expected due to the different geometry. Moreover, for the given fixed dissipation and oscillation amplitudes, transitions occur only within the window $22\, \rm{Hz} < \omega < 26\, \rm{Hz}$, which we explicitly verified for frequencies up to $85\, \rm{Hz}$. Transitions are also observed in the conservative case; however, there the angular momentum per particle dynamics are less regular than in Fig.~\ref{fig: resonant transfer}, owing to the survival of additional initial modes.

\begin{figure}
   \centering   
   \includegraphics[width=\columnwidth]{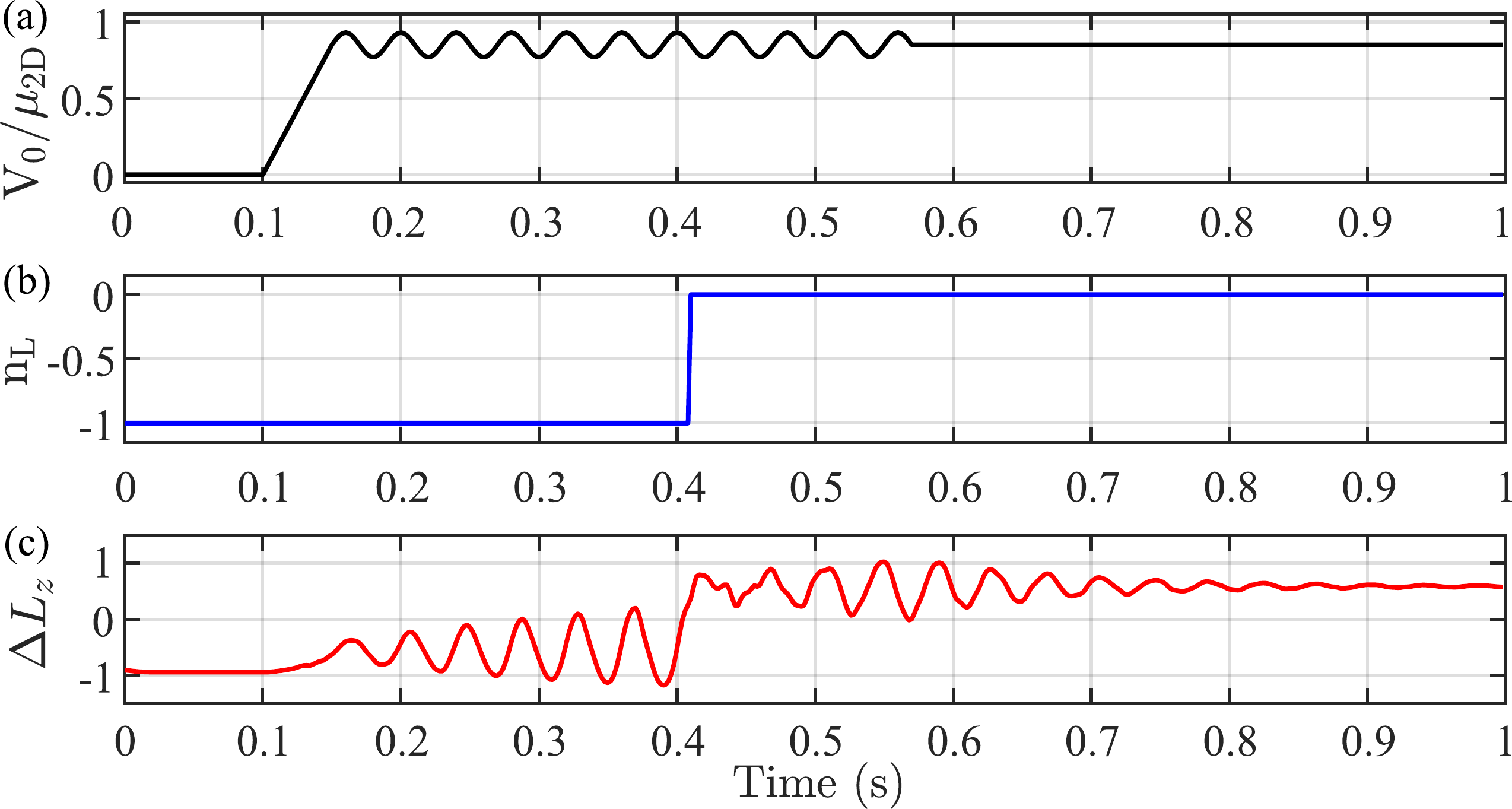}%
\caption{Example of a single resonant current transfer for temporary modulation, with dissipation $\gamma=0.001$. The initial part is identical to Fig.\ref{fig: resonant transfer}. Dynamics of (a) modulating barrier amplitude, (b) winding number in the left ring, and (c) angular momentum per particle difference.}
\label{fig: resonant transfer 2}
\end{figure}

We also note that the reported transition [Fig.~\ref{fig: resonant transfer}] indeed corresponds to angular momentum transfer. As clear evidence, Fig.~\ref{fig: resonant transfer 2} shows a protocol where we suspend modulation after the first transfer [see $t > 0.57\, \rm{s}$ in Fig.~\ref{fig: resonant transfer 2}(a)], after which the anti-vortex remains in the right ring while excitations gradually decay. Regarding the density dynamics, we further note that in these routines one can observe the vortex transfer directly, as $V_0<\mu_\text{2D}$ the density in the inter-ring region is non-zero, and the vortex core is visible.

The transfer period in Fig.~\ref{fig: resonant transfer} is roughly $\rm{0.28\, s}$, but it depends on several factors such as modulation frequency, modulation amplitude, and dissipation. The dynamics of the angular momentum difference [Fig.~\ref{fig: resonant transfer}(c)] can be qualitatively understood by analogy with the classical driven harmonic oscillator. Of course, the present case is more complex, as many modes are excited (see the hydrodynamic model of \cite{Zaremba_1998}); however, near resonance one can argue that a single mode remains dominant. Our observations under parameter variation agree with expectations: higher dissipation slows down the transfer, while larger modulation amplitude accelerates it.

Here, we have focused on identifying the resonance frequencies at which vortex transfer occurs, leaving a systematic analysis of the dependence on drive amplitude and dissipation to future work. Nevertheless, the existence of a robust resonant window provides a clear experimental target and demonstrates that phonon-mediated vortex transfer can be controllably induced by barrier modulation.

\subsection{Experimental implications}
\label{sec:exp_impl}
Our results predict experimentally accessible signatures of acoustically mediated circulation exchange in coupled rings. The required ingredients (two annular condensates with independently prepared flow states and a tunable weak link enabling controlled coupling and driving) are available with current atomtronic technology, including ring-SQUID architectures \cite{Amico2021,Ryu2013,Eckel2014,PhysRevLett.110.025302}.

A particularly robust test of persistent-current transfer is based on final-state readout: preparing a known winding, one identifies circulation exchange by measuring the \emph{final} circulation after a controlled drive, e.g., through phase-slip–mediated switching at a weak link \cite{PhysRevLett.110.025302,Ryu2013,Eckel2014}. When time-resolved dynamics are required, the population imbalance $\Delta N(t)$ can be monitored by in-situ/TOF imaging, while circulation can be tracked using minimally destructive Doppler readout \cite{kumar2016minimally}.

The persistent-current exchange mechanism investigated here is controlled, in the weak-link low-energy regime, by acoustic modes whose frequencies and damping are set by the barrier parameters and dissipation. For weak barrier modulation, periodic driving of the barrier height or position produces a resonant response when the drive matches a normal-mode frequency, yielding peaks in the spectral response of $\Delta N(t)$ and circulation proxies \cite{wang2015resonant,Gauthier2019}. Mode frequencies can be obtained from the driven response of $\Delta N(t)$, while damping rates can be extracted from the decay of the imbalance oscillations after the drive is switched off (ringdown) or from the resonance width in a frequency sweep; circulation is then determined from winding-number readout in the final state (or via repeated stroboscopic snapshots). The primary control parameters are the weak-link height and driving protocol \cite{Ryu2013,Eckel2014,wang2015resonant}, an imposed bias between the rings implemented via trap asymmetry or acceleration \cite{chaika2024acceleration}, and the effective dissipation set by temperature \cite{Meppelink2009}. Circulation exchange can be confirmed by correlated winding changes (phase slips) together with the response of $\Delta N(t)$ and circulation-sensitive diagnostics such as Doppler readout \cite{PhysRevLett.110.025302,kumar2016minimally,Ryu2013}.

\section{Conclusion}
We showed that persistent current oscillations in coupled superfluid rings originate from low-energy collective excitations with phonon-like character. These oscillations arise from angular momentum transfer mediated by sound waves propagating through the system. Our Bogoliubov-de Gennes analysis confirms that the dominant frequency corresponds to the time required for a sound mode to circulate the effective ring length, matching dynamical Gross–Pitaevskii simulations.

In the experimentally relevant regime of slow inter-ring barrier modulation, only the lowest normal mode is significantly excited, effectively setting the characteristic oscillation frequency across different interaction strengths and geometries. Such circumstances also ensure the use of the phenomenological dissipation, so in principle one can adjust the value of the phenomenological dissipation parameter $\gamma$ to match experimentally realistic decay rates. A simplified one-dimensional hydrodynamic model captures the essential features of the dynamics, including beating and damping, and yields quantitatively accurate predictions. 

The acoustic description we develop here emphasizes the role of bulk condensate physics in governing circulation exchange. This stands in contrast to our previous ghost-vortex model \cite{Bland_2022}, which was built on the notion of a point-like vortex propagating outside the Thomas–Fermi radius along the condensate boundary. That earlier framework was successful in qualitative description of persistent current oscillations' properties: their frequency and critical damping behavior. However, it relied on fine-tuning of the initial vortex position to match numerical data.
This tuning was necessary because the continuous range of ghost-vortex orbits resulted in a continuous set of values of physical observables, which are discrete for the actual system. By contrast, the present acoustic-mode approach naturally incorporates discreteness via sound-like modes in a restricted geometry. It requires no adjustable parameters, and achieves quantitative agreement with full Gross–Pitaevskii and Bogoliubov-de Gennes calculations. The success of the acoustic approach demonstrates that the oscillatory exchange of angular momentum between the rings is a collective hydrodynamic phenomenon rather than the motion of an individual vortex core.

Extending beyond previous studies \cite{Bland_2022,chaika2024acceleration}, we demonstrated that periodic modulation of the barrier at resonant frequencies enables controlled vortex transfer even for lower barrier amplitudes ($V_0<\mu_{\rm{2D}})$,
where a finite density persists in the central region of the double-ring system, through selective excitation of collective modes. Although density modulations are often emphasized in hydrodynamic analyses, our results highlight the central role of the inhomogeneous phase distribution. In the presence of quantized vorticity, it is the phase dynamics that governs circulation exchange, placing it at the core of the system's coherent evolution.

\section*{Acknowledgements}
We acknowledge Yurii Borysenko and Yuriy Bidasyuk for useful discussions. AC and AO acknowledge support from the National Research Foundation of Ukraine (Grant No. 2020.02/0032). AY is supported by PRIN Project ‘‘Quantum Atomic Mixtures: Droplets, Topological Structures, and Vortices’’. TB is supported by the Knut and Alice Wallenberg Foundation (Grant No. KAW 2018.0217) and the Swedish Research Council (Grant No. 2022-03654vr). ME is supported by U.S. National Science Foundation grant no. PHY-2207476.

\begin{appendix}
\section{Vorticity influence on the dissipative eigenspectrum}

As noted earlier, accounting for the persistent current leads to frequency splitting \eqref{eq: dispersion relation}, which in the conservative case gives $\omega_\pm = q(v_0 \pm c)$, corresponding to different group velocities of excitations in the laboratory frame. As discussed in \cite{chaika2024acceleration}, the phenomenological equations \eqref{eq:dGPE2D} implicitly assume that dissipation occurs relative to a stationary thermal cloud, resembling a classical background wind. Consequently, faster excitations are expected to decay more strongly, which is also in agreement with the Landau damping mechanism \cite{pitaevskii1997landau}.

\begin{figure}
   \centering   
   \includegraphics[width=\columnwidth]{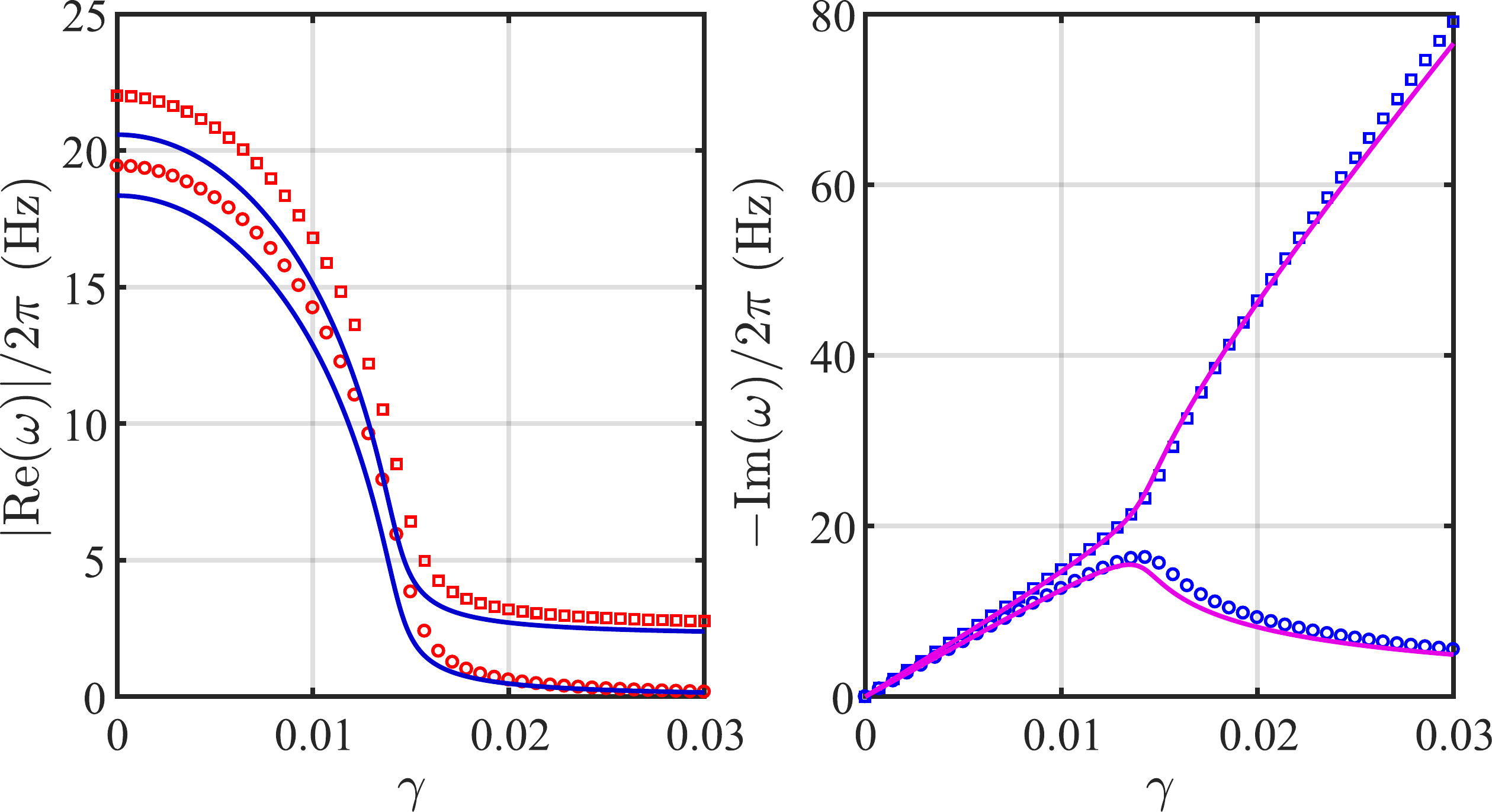}%
\caption{Excitation spectrum of real (left) and imaginary part (right) of the lowest acoustic modes for the open-gate stationary solution with four vortices in it, as a function of dissipation parameter $\gamma$. The points present the numerical solution of \eqref{eq: BdG system}, squares present in-flow excitations, circles counter-flow excitations; the lines show the acoustic model approach.}
\label{fig: dissipative eigenmodes4}
\end{figure}

In Fig.~\ref{fig: dissipative eigenmodes4} we show the excitation spectrum, similar to Fig.~\ref{fig: dissipative eigenmodes}, but restricted to the lowest acoustic pair ($n=1$) for the case of four vortices. Here, oppositely propagating modes have different decay rates, and no bifurcation appears, allowing us to distinguish between them. At low $\gamma$, their decay rates are proportional to their group velocities, resembling viscous resistance. This agrees with Ref.~\cite{pitaevskii1997landau}, where the dissipation rate was shown to be proportional to frequency in a perturbative approach. At higher $\gamma$, the decay rates approach those of the $v_0=0$ case, but both waves retain a finite real part. One finds that the initially faster mode becomes a rapidly decaying excitation with velocity $\approx 2 v_0$, while the slower mode turns into a spatially and temporally “frozen’’ mode, as both its frequency and decay rate tend toward zero with increasing $\gamma$.

The high-$\gamma$ behavior may appear puzzling; however, such values are already near the physical limits of the model parameters \cite{weiler2008spontaneous,rooney_2013,Liu18Dynamical,Ota18Collisionless,szigeti-ring}. Although $\gamma$ is a small dimensionless constant, waves with length scales larger than $\xi \gamma^{-1}$ (where $\xi$ is the healing length) enter an overdamped, non-perturbative regime. This highlights another caution: the phenomenological dissipation model should be applied carefully for very large systems.

\end{appendix}

\bibliography{References}

\end{document}